\documentstyle[twoside,psfig]{article}

\catcode`\@=11
\long\def\@makefntext#1{
\protect\noindent \hbox to 3.2pt {\hskip-.9pt  
$^{{\eightrm\@thefnmark}}$\hfil}#1\hfill}		

\def\@makefnmark{\hbox to 0pt{$^{\@thefnmark}$\hss}}	
	
\def\ps@myheadings{\let\@mkboth\@gobbletwo
\def\@oddhead{\hbox{}
\rightmark\hfil\eightrm\thepage}   
\def\@oddfoot{}\def\@evenhead{\eightrm\thepage\hfil
\leftmark\hbox{}}\def\@evenfoot{}
\def\sectionmark##1{}\def\subsectionmark##1{}}

\oddsidemargin=\evensidemargin
\newcounter{sectionc}\newcounter{subsectionc}\newcounter{subsubsectionc}
\renewcommand{\section}[1] {\vspace{12pt}\addtocounter{sectionc}{1} 
\setcounter{subsectionc}{0}\setcounter{subsubsectionc}{0}\noindent 
	{\tenbf\thesectionc. #1}\par\vspace{5pt}}
\renewcommand{\subsection}[1] {\vspace{12pt}\addtocounter{subsectionc}{1} 
	\setcounter{subsubsectionc}{0}\noindent 
	{\bf\thesectionc.\thesubsectionc. {\kern1pt \bfit #1}}\par\vspace{5pt}}
\renewcommand{\subsubsection}[1] {\vspace{12pt}\addtocounter{subsubsectionc}{1}
	\noindent{\tenrm\thesectionc.\thesubsectionc.\thesubsubsectionc.
	{\kern1pt \tenit #1}}\par\vspace{5pt}}
\newcommand{\nonumsection}[1] {\vspace{12pt}\noindent{\tenbf #1}
	\par\vspace{5pt}}

\newcounter{appendixc}
\newcounter{subappendixc}[appendixc]
\newcounter{subsubappendixc}[subappendixc]
\renewcommand{\thesubappendixc}{\Alph{appendixc}.\arabic{subappendixc}}
\renewcommand{\thesubsubappendixc}
	{\Alph{appendixc}.\arabic{subappendixc}.\arabic{subsubappendixc}}

\renewcommand{\appendix}[1] {\vspace{12pt}
        \refstepcounter{appendixc}
        \setcounter{figure}{0}
        \setcounter{table}{0}
        \setcounter{lemma}{0}
        \setcounter{theorem}{0}
        \setcounter{corollary}{0}
        \setcounter{definition}{0}
        \setcounter{equation}{0}
        \renewcommand{\thefigure}{\Alph{appendixc}.\arabic{figure}}
        \renewcommand{\thetable}{\Alph{appendixc}.\arabic{table}}
        \renewcommand{\theappendixc}{\Alph{appendixc}}
        \renewcommand{\thelemma}{\Alph{appendixc}.\arabic{lemma}}
        \renewcommand{\thetheorem}{\Alph{appendixc}.\arabic{theorem}}
        \renewcommand{\thedefinition}{\Alph{appendixc}.\arabic{definition}}
        \renewcommand{\thecorollary}{\Alph{appendixc}.\arabic{corollary}}
        \renewcommand{\theequation}{\Alph{appendixc}.\arabic{equation}}
        \noindent{\tenbf Appendix \theappendixc #1}\par\vspace{5pt}}
\newcommand{\subappendix}[1] {\vspace{12pt}
        \refstepcounter{subappendixc}
        \noindent{\bf Appendix \thesubappendixc. {\kern1pt \bfit #1}}
	\par\vspace{5pt}}
\newcommand{\subsubappendix}[1] {\vspace{12pt}
        \refstepcounter{subsubappendixc}
        \noindent{\rm Appendix \thesubsubappendixc. {\kern1pt \tenit #1}}
	\par\vspace{5pt}}

\newcommand{\textlineskip}{\baselineskip=13pt}
\newcommand{\smalllineskip}{\baselineskip=10pt}

\def\eightcirc{
\begin{picture}(0,0)
\put(4.4,1.8){\circle{6.5}}
\end{picture}}
\def\eightcopyright{\eightcirc\kern2.7pt\hbox{\eightrm c}} 

\newcommand{\copyrightheading}[1]
	{\vspace*{-2.5cm}\smalllineskip{\flushleft
	{\footnotesize International Journal of Modern Physics A, #1}\\
	{\footnotesize $\eightcopyright$\, World Scientific Publishing
	 Company}\\
	 }}

\def\abstracts#1#2#3{{
	\centering{\begin{minipage}{4.5in}\baselineskip=10pt\footnotesize
	\parindent=0pt #1\par 
	\parindent=15pt #2\par
	\parindent=15pt #3
	\end{minipage}}\par}} 

\renewenvironment{thebibliography}[1]
	{\frenchspacing
	 \ninerm\baselineskip=11pt
	 \begin{list}{\arabic{enumi}.}
	{\usecounter{enumi}\setlength{\parsep}{0pt}
	 \setlength{\leftmargin 12.7pt}{\rightmargin 0pt} 
	 \setlength{\itemsep}{0pt} \settowidth
	{\labelwidth}{#1.}\sloppy}}{\end{list}}

\newcounter{itemlistc}
\newcounter{romanlistc}
\newcounter{alphlistc}
\newcounter{arabiclistc}

\newcommand{\fcaption}[1]{
        \refstepcounter{figure}
        \setbox\@tempboxa = \hbox{\footnotesize Fig.~\thefigure. #1}
        \ifdim \wd\@tempboxa > 5in
           {\begin{center}
        \parbox{5in}{\footnotesize\smalllineskip Fig.~\thefigure. #1}
            \end{center}}
        \else
             {\begin{center}
             {\footnotesize Fig.~\thefigure. #1}
              \end{center}}
        \fi}

\newcommand{\tcaption}[1]{
        \refstepcounter{table}
        \setbox\@tempboxa = \hbox{\footnotesize Table~\thetable. #1}
        \ifdim \wd\@tempboxa > 5in
           {\begin{center}
        \parbox{5in}{\footnotesize\smalllineskip Table~\thetable. #1}
            \end{center}}
        \else
             {\begin{center}
             {\footnotesize Table~\thetable. #1}
              \end{center}}
        \fi}

\def\@citex[#1]#2{\if@filesw\immediate\write\@auxout
	{\string\citation{#2}}\fi
\def\@citea{}\@cite{\@for\@citeb:=#2\do
	{\@citea\def\@citea{,}\@ifundefined
	{b@\@citeb}{{\bf ?}\@warning
	{Citation `\@citeb' on page \thepage \space undefined}}
	{\csname b@\@citeb\endcsname}}}{#1}}

\newif\if@cghi
\def\cite{\@cghitrue\@ifnextchar [{\@tempswatrue
	\@citex}{\@tempswafalse\@citex[]}}
\def\citelow{\@cghifalse\@ifnextchar [{\@tempswatrue
	\@citex}{\@tempswafalse\@citex[]}}
\def\@cite#1#2{{$\null^{#1}$\if@tempswa\typeout
	{IJCGA warning: optional citation argument 
	ignored: `#2'} \fi}}

\def\pmb#1{\setbox0=\hbox{#1}
	\kern-.025em\copy0\kern-\wd0
	\kern.05em\copy0\kern-\wd0
	\kern-.025em\raise.0433em\box0}

\def\fnt#1#2{\footnotetext{\kern-.3em
	{$^{\mbox{\scriptsize #1}}$}{#2}}}

\def\fpage#1{\begingroup
\voffset=.3in
\thispagestyle{empty}\begin{table}[b]\centerline{\footnotesize #1}
	\end{table}\endgroup}

\def\runninghead#1#2{\pagestyle{myheadings}
\markboth{{\protect\footnotesize\it{\quad #1}}\hfill}
{\hfill{\protect\footnotesize\it{#2\quad}}}}
\font\tenrm=cmr10
\font\tenit=cmti10 
\font\tenbf=cmbx10
\font\bfit=cmbxti10 at 10pt
\font\ninerm=cmr9

\font\eightrm=cmr8

\def\qed{\hbox{${\vcenter{\vbox{			
   \hrule height 0.4pt\hbox{\vrule width 0.4pt height 6pt
   \kern5pt\vrule width 0.4pt}\hrule height 0.4pt}}}$}}

\def \missing {$E_{T} \mbox{\hspace{-0.42cm}}/ \mbox{\hspace{0.31cm}}$}

\def \misspar {$E_{T} \mbox{\hspace{-0.42cm}}/ \mbox{\hspace{0.22cm}}$}

\begin{document}

\runninghead{Search for Second and Third Generation Leptoquarks at CDF} 
            {Richard Haas, University of Florida}

\normalsize\textlineskip
\thispagestyle{empty}
\setcounter{page}{1}

\copyrightheading{}			


\vspace*{0.88truein}

\fpage{1}
\centerline{\bf SEARCH FOR SECOND AND THIRD GENERATION}
\vspace*{0.035truein}
\centerline{\bf LEPTOQUARKS AT CDF}
\vspace*{0.37truein}
\centerline{\footnotesize RICHARD HAAS\footnote{rhaas@cdfsga.fnal.gov}}
\vspace*{0.015truein}
\centerline{\footnotesize (Representing the CDF Collaboration)}
\vspace*{0.015truein}
\centerline{\footnotesize\it Department of Physics, University of Florida}
\baselineskip=10pt
\centerline{\footnotesize\it Gainesville, FL 32611, USA}

\vspace*{0.21truein}
\abstracts{
   We report the results of a search for second and third generation
   leptoquarks using 88 $\mbox{pb}^{-1}$ of data recorded by the Collider
   Detector at Fermilab. Color triplet technipions, which play
   the role of scalar leptoquarks, are investigated due to their  
   potential production in decays of strongly coupled color octet 
   technirhos. Events with a signature of two heavy flavor jets and  
   missing energy may indicate the decay of a second (third) generation 
   leptoquark to a charm (bottom) quark and a neutrino.
   As the data is found to be consistent with Standard Model
   expectations, mass limits are determined.
}{}{}

\vspace*{1pt}\textlineskip     
\section{Introduction}         
\vspace*{-0.5pt}
\noindent

The associations between quarks and leptons 
exemplified by the cancelation of triangle anomalies preserving the
renormalizability of the Standard Model
provide inviting hints at potential, fundamental connections.
Theories incorporating leptoquarks furnish a mechanism whereby quarks 
and leptons can couple directly through a Yukawa interaction of
strength $\lambda$.\cite{LQover,TC,lane_ramana}
The results of various experiments indicate that the leptoquark 
interactions should conserve baryon and lepton number and that the 
leptoquarks should couple to fermions of the same generation in order 
to suppress flavor changing neutral currents.\cite{davidson}

At the Tevatron, the principal mechanisms for pair production of
leptoquarks are $q \bar{q}$ annihilation and gluon fusion
through either direct coupling to the gluon (``continuum'') 
or a technicolor resonance state. A signature of two heavy flavor
jets, missing transverse energy (\misspar), and the absence of
leptons is utilized at CDF to search for pair produced second and
third generation leptoquarks decaying to
$c \bar{c} \nu \bar{\nu}$ and $b \bar{b} \nu \bar{\nu}$,
respectively.\cite{LQmine}
Events are required to have two or three jets with 
$E_{T} \geq 15$ GeV, \missing $\geq 40$ GeV, jets well separated in
$\phi$ from both the direction of \missing and from each other, and no
leptons. Heavy flavor jets are identified through a jet probability
algorithm.\cite{jetprob}

\vspace*{1pt}\textlineskip
\section{Continuum Leptoquarks}
\vspace*{-0.5pt}
\noindent

The pair production cross section for scalar leptoquarks is parameter 
free and known to next-to-leading order.\cite{kraemer} 
Vector leptoquark interactions include the model dependent 
couplings $\kappa_{G}$ and $\lambda_{G}$.\cite{blum}
Yang-Mills type coupling ($\kappa_{G} = \lambda_{G} = 0$)
and minimal coupling ($\kappa_{G} = 1$ and $\lambda_{G} = 0$) are
investigated.
At present only leading order processes have been calculated for
vector leptoquark pair production.\cite{blum}

\begin{figure}
   \vspace*{13pt}
   \centerline{\psfig{file=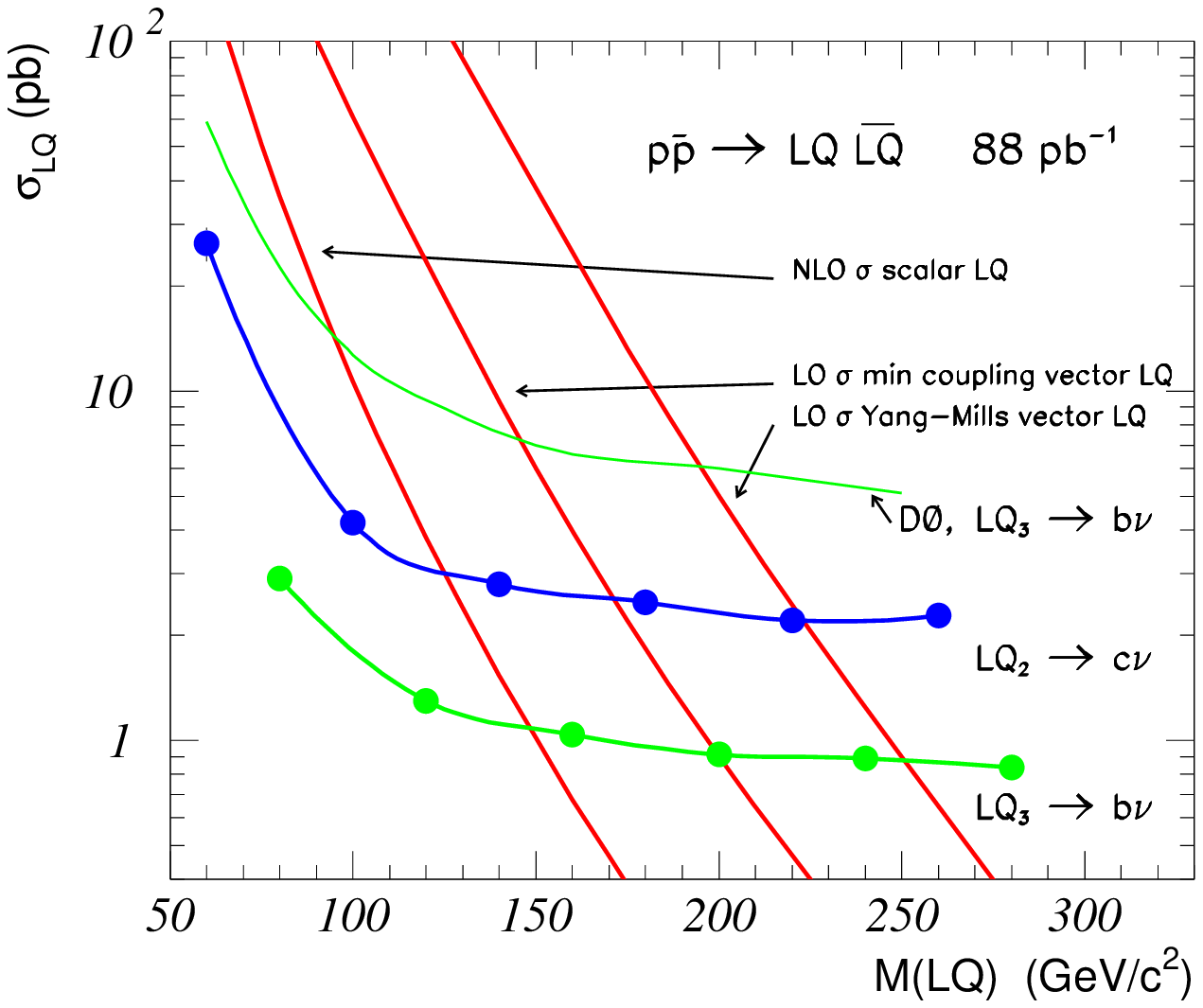,width=2.8in}
               \psfig{file=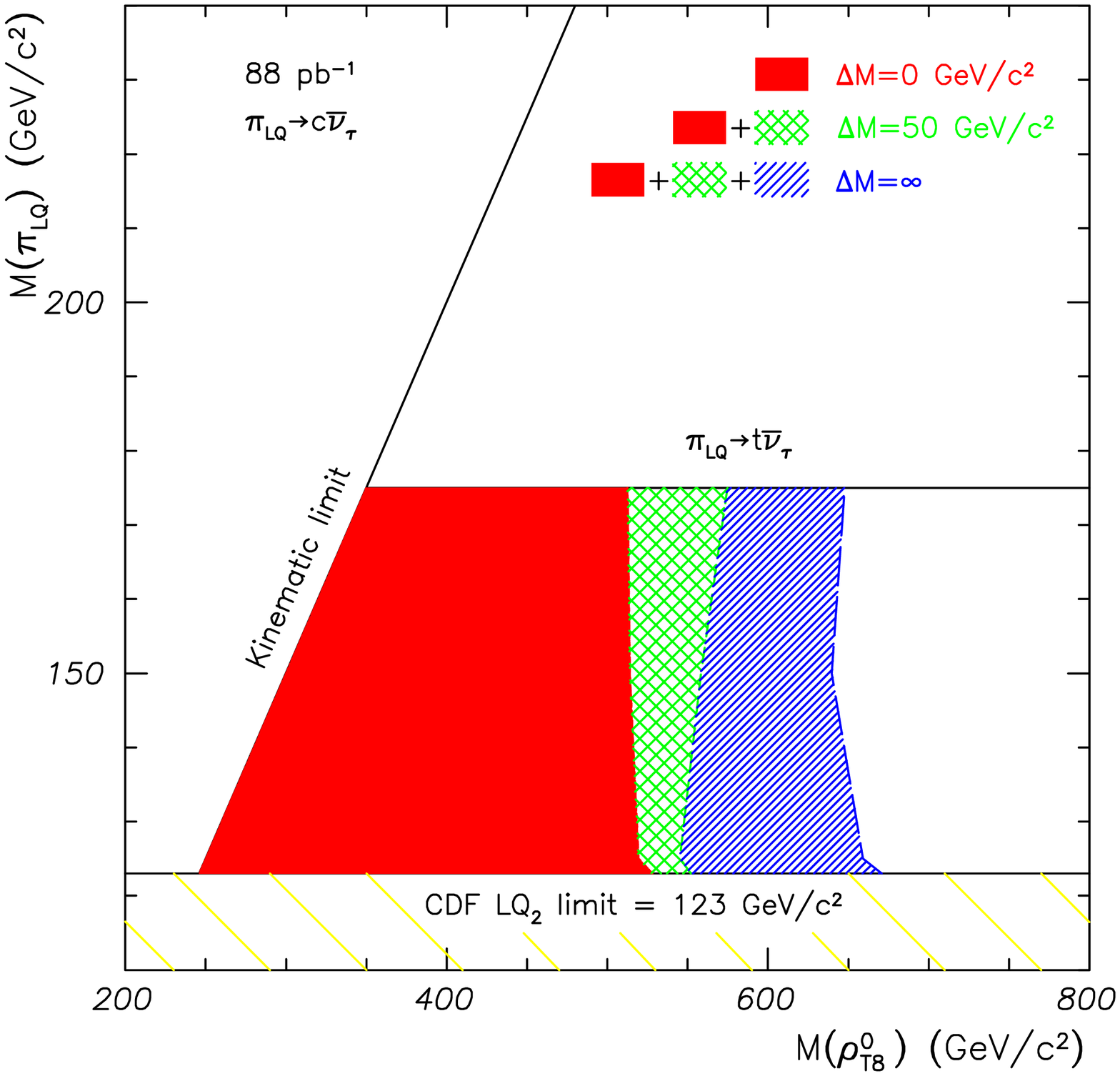,width=2.4in}}
   \vspace*{13pt}
   \fcaption{Left plot: 95\% CL cross section limit on 
             $LQ_{2} \rightarrow \nu_{\mu} c$ and on 
             $LQ_{3} \rightarrow \nu_{\tau} b$ together with the 
             expected theoretical cross sections for scalar and two
             types of vector LQs.
             Right plot: The 95\% CL limit for the process 
             $\rho_{T8} \rightarrow \pi_{LQ} \bar{\pi}_{LQ}$ in which 
             $\pi_{LQ} \rightarrow c \bar{\nu}_{\tau}$.
            }
   \label{fig:CLcontin}
\end{figure}

The 95\% CL limits for continuum leptoquark production cross
sections are determined and compared to the
corresponding theoretical cross sections.\cite{kraemer,blum}
The results are shown in the left plot of Figure \ref{fig:CLcontin}.
In the case of second generation leptoquarks, 
scalar leptoquarks with $M < 123$ GeV/c$^{2}$, 
minimally coupled vector leptoquarks with $M < 171$ GeV/c$^{2}$, and 
Yang-Mills vector leptoquarks with $M < 222$ GeV/c$^{2}$ are excluded.
For third generation leptoquarks,
scalar leptoquarks with $M < 148$ GeV/c$^{2}$,
minimally coupled vector leptoquarks with $M < 199$ GeV/c$^{2}$, and 
Yang-Mills vector leptoquarks with $M < 250$ GeV/c$^{2}$ are excluded.

\vspace*{1pt}\textlineskip
\section{Leptoquarks from Technicolor}
\vspace*{-0.5pt}
\noindent

Enhancement of leptoquark pair production may occur through the 
decay of technicolor resonance states.\cite{TC,lane_ramana}
Color octet technirhos, $\rho_{T8}$, with the same quantum numbers as
the gluon are possible, allowing for $s$-channel coupling.
The color triplet and octet technipions, denoted by
$\pi_{LQ}$ and $\pi_{T8}$, couple in a 
Higgs-like fashion to quarks and leptons with the $\pi_{LQ}$ 
identified as a scalar leptoquark. 
The leading-order cross section is sensitive to 
$\Delta M = M(\pi_{T8}) - M(\pi_{LQ})$, $\Delta M = 50$
GeV/c$^{2}$ expected.\cite{lane_ramana}

\begin{figure}
   \vspace*{13pt}
   \centerline{\psfig{file=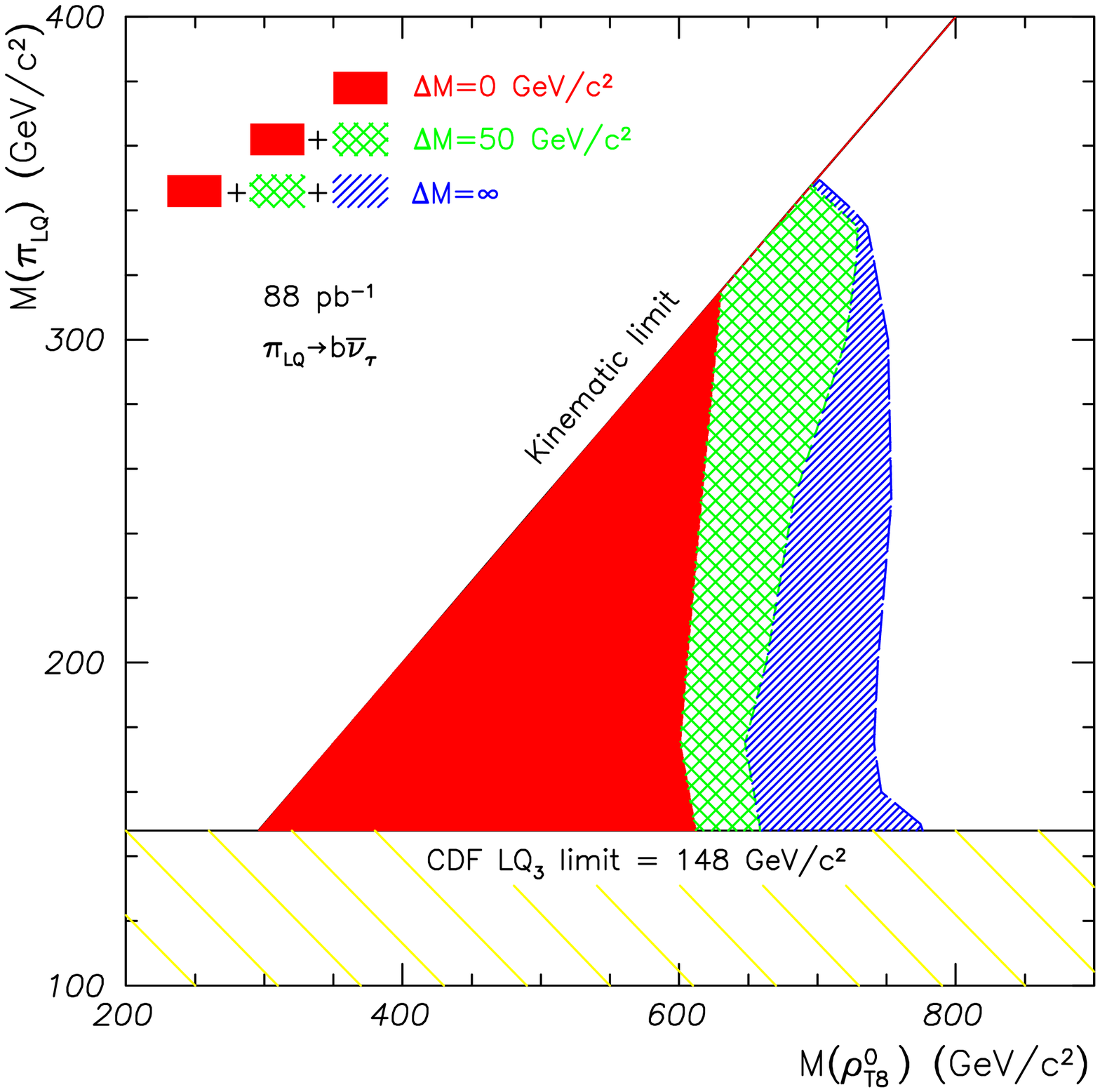,width=2.4in}
               \hspace{0.2in}
               \psfig{file=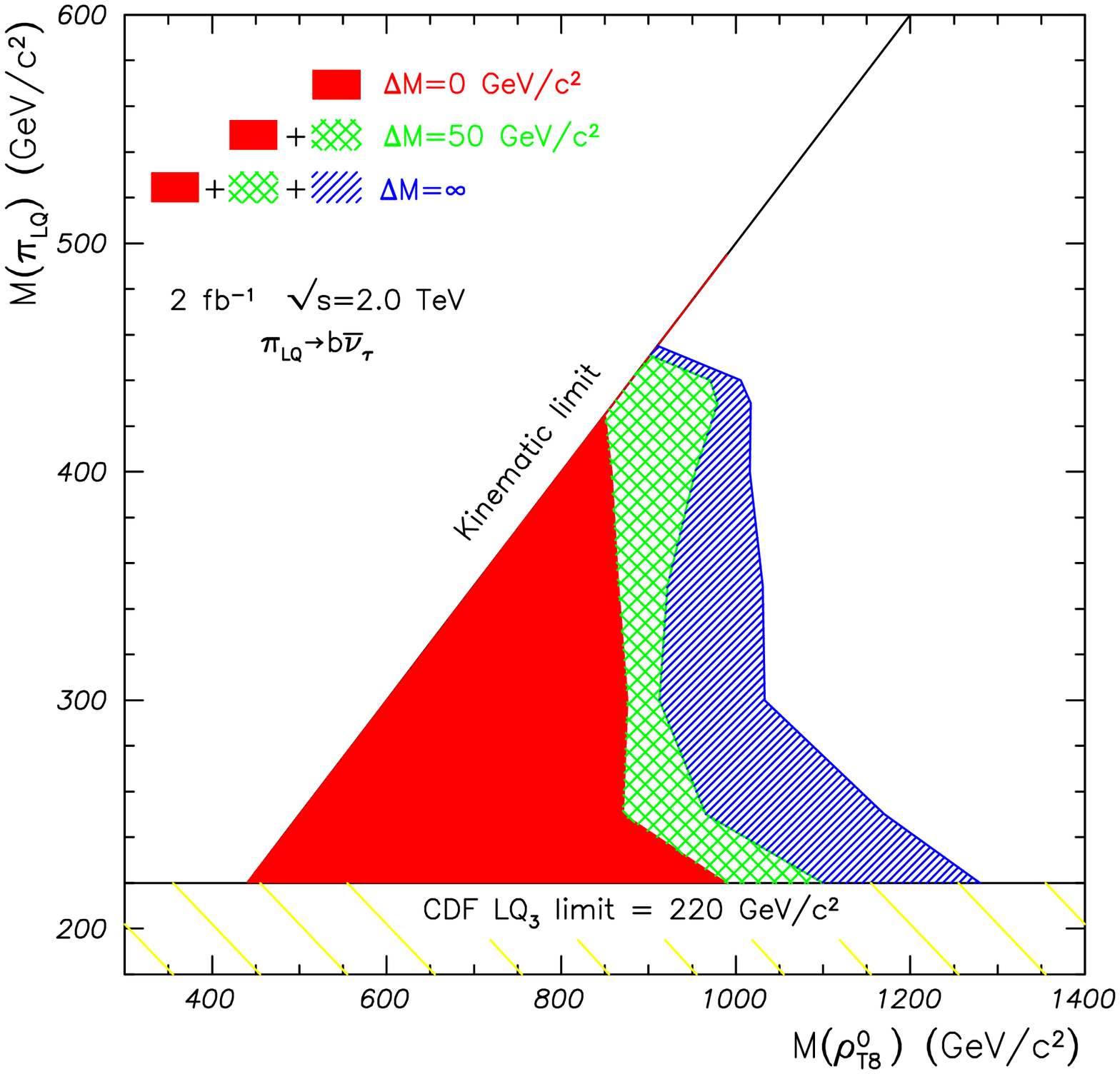,width=2.4in}}
   \vspace*{13pt}
   \fcaption{Left plot: The 95\% CL limit for the process 
             $\rho_{T8} \rightarrow \pi_{LQ} \bar{\pi}_{LQ}$ in which 
             $\pi_{LQ} \rightarrow b \bar{\nu}_{\tau}$.
             Right plot: Run II projection for the third generation
             continuum and resonantly produced leptoquarks.
            }
   \label{fig:CLtech}
\end{figure}

The 95\% CL exclusion regions in the $M(\rho_{T8}) - M(\pi_{LQ})$
plane for $\Delta M =0$, $50$ GeV/c$^{2}$, and $\infty$  
are shown as shaded areas in the right plot of 
Figure \ref{fig:CLcontin} and the left plot of Figure \ref{fig:CLtech}.
The kinematically forbidden region is given by 
$M(\rho_{T8}) < 2 M (\pi_{LQ})$. 
In the right plot of Figure \ref{fig:CLcontin}, the decay of the 
leptoquark to $c \bar{\nu}_{\tau}$ is limited by the top quark mass, 
above which the leptoquark will decay preferentially to $t \bar{\nu}_{\tau}$. 
When $\Delta M = 0$, $M(\rho_{T8}) < 510$ GeV/c$^{2}$ for the second
generation and $M(\rho_{T8}) < 600$ GeV/c$^{2}$ for the third
generation are excluded at 95\% CL.

\vspace*{1pt}\textlineskip
\section{Run II Projections}
\vspace*{-0.5pt}
\noindent

The anticipated start of the Tevatron's Run II in March 2001 will
provide fertile ground for further searches of new phenomena. 
With a significantly improved detector, an increase in the center of
mass energy to 2.0 TeV, and a projected integrated luminosity of 
2 $\mbox{fb}^{-1}$ by 2003, the sensitivity to continuum produced 
third generation leptoquarks extends to $M < 220$ GeV/c$^{2}$. For the
leptoquarks produced from technirho decays, the Run II 95\% CL
exclusion region is shown in the right plot of Figure \ref{fig:CLtech}. 
The sensitivity to $\rho_{T8}$ extends to approximately 1 TeV.


\nonumsection{References}

\end{document}